\documentstyle[11pt]{article}
\newcommand{\rinv}{\stackrel{\rightarrow}{G_{(0)}^{-1}}}
\newcommand{\linv}{\stackrel{\leftarrow}{G_{(0)}^{-1}}}

\begin{document}

\begin{tabbing}
\`SUNY-NTG-92-11\\
\`April 1992
\end{tabbing}
\vbox to  0.8in{}
\centerline{\Large \bf  Nonequilibrium Quantum Field Kinetics }
\vskip 2.5cm
\centerline{\large A.  Makhlin }
\vskip .3cm
\centerline{Department of Physics}
\centerline{State University of New York at
Stony Brook}
\centerline{Stony Brook, New York 11794}
\vskip 0.35in
\centerline{\bf Abstract}
\indent
 Using the general framework of nonequilibrium statistical mechanics for
relativistic quantum field systems we derive the basic equations of
quantum field kinetics. The main  aim of the approach is calculation of
observables associated with quark-gluon plasma which is out of thermal and
chemical equilibrium. We show that in the regime of high rate of the phase and
statistical mixing  the perturbation theory for many-body quantum field system
should have the form which much differs from the standard expansion in powers
of
coupling constant.

\vskip .25cm
\vfil
${^\dagger}$ E-mail address: "makhlin@sbnuc"
\noindent

\eject
\newpage
\pagestyle{plain}
\addtocounter{page}{-1}

{\bf \Large 1.Introduction}

The thermal quantum field theory have got an intensive development in the last
decades. There is a set of problems concerning
exotic state of the matter created in  early Universe or heavy ions
collisions where its relativistic version is insistently demanded.
The temperature field theory is usually considered to be a formalism which
allows one to account for the temperature and density influence at the level
without interaction and keep the perturbation theory in the form of Feynman
diagram technique.

The first one and the most often used was the imaginary-time formalism of
Matsubara [1] given later by Feynman in the path integral form [2]. It may
me reduced to the imposing on the propagators of free wave equations the
periodic boundary conditions in the imaginary direction of complex time
$\;t\;$ [3]. The theory is assumed to describe the thermal equilibrium
after $\;{\rm Re}t\;$ is eliminated.
Then it comes to be the Euclidian field theory.
The back way to the physical region of real time goes through an analytic
continuation which is not well justified even for nonrelativistic systems
[4].

Another approach, the thermofield dynamics (TFD) had been elaborated by
Umesawa and Takahashi [5]. This formalism introduces ghosts in such a way
that the average over the mixed state of Gibbs ensemble gets the formal
shape of the average over a pure state (the so-called thermal vacuum) in
an extended Hilbert space. The physical meaning of this ghost fields
is not so clear and the approach doesn't contain a unique prescription
for calculation of the observables.

One version more was designed by Keldysh [6] especially for nonequilibrium
systems.  Though in the most of recent papers (e.g.[7]) it is considered
as one of the versions of the TFD, specified by a certain choice of the path
in the plane of complex time, it is a quite separate approach. It contains no
thermal doublets with the ghost components. Its matrix structure is completely
physical because this technique corresponds to the Heisenberg picture and do
not
assume that the ground state is stable. It is applicable to thermal equilibrium
as well and describes its kinetic background. This technique introduces the
heat
ensemble (on the same ground as any other ensemble) via the density matrix of
 initial state. The time evolution of the system is governed by the
usual equations of motion for the Heisenberg field operators. All
the observables are unambiguously defined from the first principles. Moreover,
the technique creates itself in course of calculation of the Heisenberg
observables at given density matrix of the initial state.

 A goal of this paper  and the next  three
papers [8,9,10] is to calculate the rate of dileptons and photons emission
from quark-gluon plasma.  An overview of many publications on this subject
revealed that different authors even were started from the different
definitions
for the very emission rate [11-13].  All previous papers considered  QGP as
a totally  equilibrium object and essentially used the detail balance
relationships. Some very restrictive statements with a strength of a theorem
(like thermal version of the KLN theorem) was proved in these frames [12].
The hard loops resummation technique [14] and a proof of the gauge
independence of a plasmon damping are essentially based on the idea of a
thermal equilibrium also.

  The later theoretical studies revealed that quark-gluon plasma hardly
reach true thermal and chemical equilibrium [15]. This conclusion was
confirmed by a various computer simulations [16]. So all the references to
a detail balance come to be irrelevant.

  This paper presents the most general framework of the approach named
"quantum  field kinetics" (QFK). It starts from the basic definition of
observables like photon or dilepton rate of emission from the QGP in terms
of their Heisenberg operators and the density matrix of an arbitrary  initial
state of the emitting plasma. Then we step by step build the technique for
calculation of namely these quantities.

We show that the technique proposed by Keldysh [6] for nonequilibrium systems
is most suitable for this purpose and re-derive all the basic equations for
the case of relativistic QGP.  We do not study any phase-space distributions
and do not try to get any kinetic equation of standard form. Instead we show
that any observable
may be split into the parts separately contributed  by the initial data and
the current dynamics. This splitting has allowed to single out the two main
kinetic regimes of the nuclear matter, the regimes of weak and strong kinetic
coupling.

We show that the perturbative expansions in this regimes is very different
though this difference appears as the result of different ways of resummation
of
the standard perturbation series.

At the end of the paper we  discuss the
renormalization procedure. We carefully examine this point and show that
even at finite temperature renormalization can result only in {\it temperature-
independent} quasi-local counter-terms.

 \renewcommand{\theequation}{2. \arabic{equation}}
\setcounter{equation}{0}
\bigskip
{\bf \Large 2.Signals from quark-gluon plasma}
\bigskip

   We are going to design a technique which would allow one to calculate
the observables associated with QGP. For example, these are the
electromagnetic signals emitted from the hot nuclear matter created
in the A-A collisions. They are unambiguously defined at both
theoretical and apparatus level and it seems reasonable to try to
follow the history of their origin using only the basic principles
of quantum mechanics as long as it is possible. Being a success,
we'll express the signals in terms of the emitting system
parameters.

   The emitting system, the QGP, is expected to be described by standard QCD
Lagrangian and the density matrix  $\rho_{QCD}$ of the initial state of
 the QGP. We shall specify it later.

   Let us choose the inclusive cross-sections of photon and dilepton emission
as the observables. Then we must include the interaction of charged quarks,
$q(x)$, and leptons,  $\psi(x)$,  with the photon field,  $A^{\mu} (x)$,
to the total interaction Lagrangian:
\begin{eqnarray}
   {\cal L}_{int}(x) &=&
   e \bar{\psi} (x) \gamma^{\mu}\psi(x)A_{\mu}(x)
    +  e \sum_{q,i}\bar{q}_{i}(x)\gamma^{\mu}
   A_{\mu}(x)q_{i}(x) +
   +g_{0} \sum_{q,i}\bar{q}_{i}(x)t^{a}_{ij}\gamma^{\mu}
   B^a_{\mu}(x)q_{j}(x) +  & \nonumber\\
   &+& g_{0}f_{abc}\partial^{\mu}B_{a}^{\nu}(x)
   B_{\mu}^{b}(x)B_{\nu}^{c}(x)
   + (g_{0}^{2}/4)f_{abc}f_{agh}B_{b}^{\mu}(x)
   B_{c}^{\nu}(x)B_{\mu}^{g}(x)
   B_{\nu}^{h}(x)
   \end{eqnarray}

The notations need no comments. The density operator gains the projector on
vacuum initial state of photons and leptons,
\begin{equation}
  \rho = \rho_{QCD}\otimes|0_{e\gamma}\rangle\langle0_{e\gamma}|,
\end{equation}
and we shall assume that QGP remains transparent  for photons and leptons
allover its history. The $\rho_{QCD}$ is formed due to strong
interactions only.

   The Lagrangian (2.1) gives rise to the ordinary S-matrix in the
$in$-interaction picture,
\begin{equation}
S = T \exp\{i\int d^4x{\cal L}_{int}(x)\}
\end{equation}
which is supposed to be a certain limit of the evolution operator which governs
the dynamics of Heisenberg observables.

   The amplitudes for the transition from one of the initial states,
$|in\rangle$,
to the final ones containing photon or dilepton look like
\begin{equation}
 \langle X|c({\bf k},\lambda)S|in\rangle   \;\;\;   {\rm or} \;\;\;
 \langle X|b(2)a(1)S|in\rangle
\end{equation}
where
$c({\bf k},\lambda)$ and $a(J)=a({\bf k}_{J},\sigma_{J})$ and $b(J)$
are the photon, electron and positron annihilation operators. Summarizing the
squared moduli of these amplitudes over a complete set of uncontrolled states
$|X\rangle$
and averaging over the initial ensemble we find the inclusive spectra of
photons and dileptons,
\begin{eqnarray}
{  {dN_{\gamma}} \over {d{\bf k}} } &
 = \sum_{\lambda} Sp \rho_{in}
          S^{+}c^{+}({\bf k},\lambda)c({\bf k},\lambda)S, \\
{   {dN_{e^{+}e^{-}}}
                \over {d{\bf k}_{1}d{\bf k}_{2}}  }  &
 =   \sum_{\sigma_{1},\sigma_{2}}  Sp  \rho_{in}
    S^{+}a^{+}(1)b^{+}(2)b(2)a(1)S.
\end{eqnarray}

   It is easy now to commutate the Fock operators with $S$ and $S^{+}$.
Then Eq.(2.5) takes the the form
\begin{equation}
   k^{0}{{dN_{\gamma}}\over{d{\bf k}d^{4}x}} =
   {{ig_{\mu\nu}}\over{2(2\pi)^{3}}} \pi^{\mu\nu}_{10}(-k),
\end{equation}
where
\begin{equation}
 \pi^{\mu\nu}_{10}(-k)=-i\int d^{4}(x-y)e^{-ik(x-y)}
\langle{{\delta S^{+}}\over{\delta A^{\mu}(x)}}
        {{\delta S}\over{\delta A^{\nu}(y)}}\rangle
\end{equation}
is the Fourier transform of the two Heisenberg currents product averaged with
$\rho_{QCD}$,
\begin{eqnarray}
 \pi^{\mu\nu}_{10}(x,y) & =i\langle{ \delta S^{+} \over
   \delta A^{\mu}(x) }
        {{\delta S}\over{\delta A^{\nu}(y)}}\rangle=
        i\langle {\bf j}^{\mu}(x) {\bf j}^{\nu}(y)\rangle,  \\
 {\bf j}^{\mu}(x) & = S^{+}T(j^{\mu}(x)S)\equiv T^{+}(j^{\mu}(x)S^{+})S, \\
     j^{\mu}(x) & =(1/2)\sum_{i,q} e_{q}[\bar{q}_{i}(x)\gamma^{\mu},q_i (x)].
\end{eqnarray}
The dilepton rate of emission (2.6) takes the form
\begin{equation}
 k_{1}^{0} k_{2}^{0} {{dN_{e^{+}e^{-}}}\over{d{\bf k}_{1}
{d\bf k}_{2}d^{4}x}}=
 -ie^{2}{{L_{\mu\nu}(k_{1},k_{2})}\over{4(2\pi)^{6}}}
{\bf \Delta}^{\mu\nu}_{10}(-k),
  \end{equation}
where
$k=k_{1}+k_{2}$, $L^{\mu\nu}= k_{1}^{\mu} k_{2}^{\nu}+
k_{2}^{\mu} k_{1}^{\nu}-g^{\mu\nu}(k_{1}k_{2}-m_{e}^{2}) $,
is a trace of lepton spinors and
\begin{equation}
   {\bf \Delta}^{\mu\nu}_{10}(-k)=-i\int d^{4}(x-y)
  \langle T^{+}(A^{\mu}(x)S^{+}) T(A^{\mu}(y)S)\rangle
   e^{-ik(x-y)},
\end{equation}
is a kind of photon Wightman function averaged with
$\rho^{QCD}$.
The operators $A(x)$ of $in$-interaction picture and Heisenberg operators
${\bf A}(x)$ are connected via relation similar to (2.10).

   In both (2.7)and (2.12) we have assumed that an explicit separation of
long-range and short-range orders  takes place  and have passed from the
inclusive cross-sections to the emission rates per unit volume.

   We have expressed the observable rates of emission in terms of Heisenberg
operators product taken in a quite definite order. It is high time now
to choose  the formalism which would allow one to calculate namely these
quantities. We are to keep in mind that all the operators are driven by
${\cal L}_{int} $ and all the information about initial state of the system is
hidden in $\rho_{in}$.  Nothing else is needed to solve the problem.

\renewcommand{\theequation}{3. \arabic{equation}}
\setcounter{equation}{0}
\bigskip
{\bf \Large 3. Equations of relativistic quantum field kinetics.} \\
\bigskip

{\underline{\it 3.1. Basic definitions.}}
\bigskip

   The design of the technique suited for calculation of observables like
emission rate had been initiated by Keldysh[6]. It is based on a specific set
of
exact (dressed) Greenians which are nothing but the Heisenberg operators
products averaged with the density matrix of initial state. For the quark field
they look like
\begin{eqnarray}
{\bf G}_{10}(x,y) & =-i\langle{\bf q}(x)\bar{{\bf q}}(y)\rangle ,  \;\;\;
{\bf G}_{01}(x,y) & = i\langle\bar{{\bf  q}}(y){\bf q}(x)\rangle, \nonumber \\
{\bf G}_{00}(x,y) & =-i\langle T({\bf q}(x)\bar{{\bf q}}(y))\rangle ,\;\;
{\bf G}_{11}(x,y) & =-i\langle T^{+}({\bf q}(x)\bar{{\bf q}}(y))\rangle ,
 \end{eqnarray}
where $T$ and $T^{+}$
are symbols of time and anti-time orderings. They may be written in a unified
form,
\begin{equation}
   {\bf G}_{AB}(x,y)=-i\langle T_{c}({\bf q}(x_{A})
                       \bar{{\bf q}}(y_{B}))\rangle,
\end {equation}
in terms of special ordering $T_{c}$ along a contour $C=C_{0}+C_{1}$,
the doubled time axis, with $T$-ordering on $C_{0}$ and $T^{+}$- ordering on
$C_{1}$. The operators labelled by '1' are $T^{+}$-ordered
and stand before the $T$-ordered operators labelled by '0'. Recalling that
\begin{equation}
 {\bf q}(x)= S^{+}T(q(x)S)\equiv T^{+}(q(x)S^{+})S,
\end{equation}
we may introduce the formal operator $S_{c}=S^{+}S$
and rewrite (3.2) using the $in$-interaction picture
\begin {equation}
{\bf G}_{AB}(x,y)=-i\langle T_{c}(q(x_{A})\bar{q}(y_{B})S_{c})\rangle,
\end{equation}
where by the definition the internal variables of $S$ lie on
$C_{0}$ and that of  $S^{+}$ on $C_{1}$.

   The boson Greenians are built in the same manner, i.e. for gluon field,
$B(x)$, and photon field, $A(x)$,
\begin{eqnarray}
{\bf D}_{AB}(x,y)=-i\langle T_{c}(B(x_{A})B(y_{B})S_{c})\rangle,  \\
{\bf \Delta}_{AB}(x,y)=-i\langle T_{c}(A(x_{A})A(y_{B})S_{c})\rangle,
\end{eqnarray}
where the vector and colour indices are suppressed.

\bigskip
{\underline{\it 3.2. Density matrix of the initial state.}}
\bigskip

   The choice of density matrix is that of phenomena to be studied. If the
the initial state of a system is a few excitations above the perturbative
vacuum we get density matrix of a pure state. In this case we deal with the
well known picture of scattering. This situation is described by a certain set
of vacuum expectations and the density matrix is
\begin{equation}
 \rho_{QCD}=|0_{QCD}\rangle\langle 0_{QCD}|.
\end{equation}

The density matrix which we will use for explicit calculations of the rates of
the photon and dilepton emission is in line with the following scenario of the
heavy ions collision. The initial stage of a collision at RHIC or LHC energies
($\tau \sim 0.5 fm$) is a region of nucleons fragmentation and development of
the initial parton cascade. Dileptons are emitted only due to the hard
Drell-Yan
process. Our region begins a little bit later when partons are already
chaotized
and may be described by the one-particle distributions.

The most general density matrix which simulate any given ahead form
of the one-particle distribution is of the next form,
\begin{equation}
\rho = \prod_{N}\prod_{p,j}e^{-f_{j}(N,p)a^{+}_{j}(N,p) a_{j}(N,p)},
\end{equation}
where $ N $ labels the space cells on the hypersurface of the initial data
and $n_{j}(N,p)=a^{+}_{j}(N,p) a_{j}(N,p) $ is an operator of the number of
partons of the sort $j$ and quantum numbers $p$ in the $N$-th cell. Thus
we completely neglect all the correlation effects in the phase space of the
initial partons.

 Density matrix of  the Gibbs ensemble of noninteracting quarks and gluons
against the hydrodynamic background is of the same kind as (3.8) and
obey  an additional condition of the entropy extremum. Introducing the local
4-velocity of continuous media $u^{\mu}(x)$ we can write,
\begin{equation}
    \rho^{QCD}=\prod_{N}  \frac{\exp [(-P_{N}u_{N}+\mu_{N}Q_{N})
        /T_{N}]}{Sp[\exp [(-P_{N}u_{N}+\mu_{N}Q_{N})/T_{N}]]}  ,
\end{equation}
where  $P^{\mu}_{N}$ and $Q_{N}$
are the total 4-momentum and (barionic) charge of free quarks and gluons at
temperature $T_{N}$ and chemical potential $\mu_{N}$ in the small 3-volume
$V_{N}$ on the hypersurface of initial data.

   The explicit form of "free" Greenians which make the basis of perturbation
theory is quite evident. The "vacuum" Green functions are of standard form:
\begin{eqnarray}
 G^{(0)}_{10,01}(s) & =
      -2 \pi i (\hat{s}+m) \theta (\pm s_{0})\delta (s^{2}-m^{2}),\;\;\;
 D^{(0)\mu \nu}_{10,01}(s) & =
      -2 \pi i d^{\mu\nu}(s)
\theta (\pm s_{0})\delta (s^{2}) , \nonumber \\
    G^{(0)}_{00,11}(s) & = \pm \frac{\hat{s}+m}{s^{2}-m^{2} \pm i0},\;\;\;
   D^{(0)\mu \nu}_{00,11}(s)  & = \pm \frac{d^{\mu\nu}}{s^{2}\pm i0}.
\end{eqnarray}
The projector $d^{\mu\nu}(s)$  depends upon the choice of the gauge.

   The bare Greenians of any ensemble are as follows
\begin{eqnarray}
  G_{AB}(s) = G^{0}_{AB}(s)+G_{\beta}(s), \;\;\;
   D^{\mu\nu}_{AB}(s) = D^{(0)\mu\nu}_{AB}(s)+D^{\mu\nu}_{\beta}(s),
\end{eqnarray}
 The additional terms originating from the $\rho_{in}$ are
\begin{eqnarray}
     G_{\beta}(s) & = 2 \pi i (\hat{s}+m)\delta(s^{2}-m^{2})
     [\theta (s_{0})n^{(+)}(s)+
\theta (-s_{0})n^{(-)}(s)],  \nonumber   \\
     D^{\mu\nu}_{\beta}(s) & = -2 \pi i d^{\mu\nu}(s) \delta(s^{2})
     [\theta (s_{0})f^{(+)}(s)+\theta (-s_{0})f^{(-)}(s)],\;\;\;
   \end{eqnarray}
with Fermi- and Bose-occupation numbers $ n^{\pm}$ and  $ f^{\pm}$
specified by a partial choice of the density matrix.

   All the theorems of Wick type which are necessary for the theoretical and
perturbative calculations can be easily proved for the density matrices
 (3.7) and (3.8).

\bigskip
{\underline{\it 3.3. Schwinger-Dyson equations for QFK.}}
\bigskip

   Except for the matrix form, the Schwinger-Dyson equations  for Heisenberg
Greenians remain to be the same as in any other technique,
\begin{equation}
  {\bf G}_{AB} = G_{AB}+ \sum_{RS} G_{AR}\circ M_{RS}\circ {\bf G}_{SB},
\end{equation}
where the dot stands for convolution in coordinate space and for the usual
product in the momentum one (providing the system can be treated as homogeneous
in space and time). The quark self-energy matrix looks like
\begin{eqnarray}
  M_{AB}(x,y)=i(-1)^{A+B}g_{0}^{2}\sum_{R,S=0}^{1} (-1)^{R+S}
   \int d \xi d \eta \times           \\
  \times t^{a} \gamma^{\mu} {\bf G}_{AR}(x,\xi)
   \Gamma^{d,\lambda}_{RB,S}(\xi,y;\eta)
 {\bf D}^{da}_{SA,\lambda\mu}(\eta,x),     \nonumber
\end{eqnarray}
and contains the strong $qqB$-vertex,
\begin{eqnarray}
 \Gamma^{d,\lambda}_{SQ,P}(x,y;z)=(-1)^{P+S+Q}{
 {\delta [{\bf G}^{-1}(x,y)]_{SQ} } \over
 {g_{0} \delta {\cal B}^{d}_{\lambda}(z_{P}) }  }   ,
\end{eqnarray}
the functional derivative with respect to "external" field ${\cal B}(x)$.

   The photon Greenians obey a similar equations,
\begin{equation}
  {\bf \Delta}_{AB} = \Delta_{AB}+ \sum_{RS}
 \Delta_{AR}\circ \Pi_{RS}\circ {\bf \Delta}_{SB},
\end{equation}
with electromagnetic polarization operator
\begin{equation}
  \Pi_{AB}(x,y)=-i(-1)^{A+B}g_{0}^{2} \sum_{R,S=0}^{1}(-1)^{R+S}
\! \int \! d \xi d \eta \gamma^{\mu}{\bf G}_{AR}(x,\xi)
 E^{\nu}_{RS,B}(\xi,\eta;y) {\bf G}_{SA}(\eta,x),
\end{equation}
and electromagnetic vertex dressed by the strong interaction,
\begin{equation}
    E^{\lambda}_{RS,P}(x,y;z)=(-1)^{R+S+P}{
 {\delta [{\bf G}^{-1}(x,y)]_{RS} } \over
  {e \delta {\cal A}_{\lambda}(z_{P}) }  } ,
\end{equation}
which in its turn obey an equation
\begin{equation}
  E^{\mu}=\gamma^{\mu}+E^{\mu}\circ {\bf G G}\circ {\bf K},
\end{equation}
with the four-fermion vertex  ${\bf K}$.

   The gluon Greenians obey the same equations,
\begin{equation}
{\bf D}_{AB}=D_{AB}+ \sum_{RS} D_{AR}\circ {\cal P}_{RS}\circ {\bf D}_{SB},
\end{equation}
where the gluon self-energy
${\cal P}$
contains several types of vertices and loops.

   There exists a set of obvious relations
\begin{equation}
G_{00}+G_{11}=G_{10}+G_{01},\;\;\; \; M_{00}+M_{11}=-M_{10}-M_{01},
\end{equation}
as well as their copies for boson Greenians and self-energies. They indicate
that only three elements of  $2 \times 2$ matrices  $G,M$,  etc. are
independent. To remove the overdetermination let us introduce the new
functions,
\begin{eqnarray}
 G_{ret}  =G_{00}-G_{01},\;\;\; G_{adv}  =G_{00}-G_{10},\;\;\;
 G_{1}  =G_{00}+G_{11};  \nonumber \\
 M_{ret}  =M_{00}+M_{01},\;\;\; M_{adv}  =M_{00}+M_{10}, \;\;\;
 M_{1}  =M_{00}+M_{11},
\end{eqnarray}
as well as their analogs for bosons. We can pass to this reduced set by
a unitary transformation [6] ,
\begin{eqnarray}
\tilde{G}=R^{-1}GR, \; \tilde{M}=R^{-1}MR ,\;
R={{1}\over {\sqrt{2}}} \left| \begin{array}{rc}
                                            1 & 1 \\
                                           -1 & 1  \end{array} \right|.
\end{eqnarray}
In this new representation
\begin{eqnarray}
  \tilde{G}= \left| \begin{array}{ll}
                                       0 & G_{adv} \\
                                 G_{ret} & G_{1}     \end{array} \right| ,
\;  \tilde{M}= \left| \begin{array}{ll}
                                 M_{1}    & M_{ret}  \\
                                 M_{adv}  & 0      \end{array} \right|
\end{eqnarray}
Using (3.24) we can rewrite the Schwinger-Dyson equations (3.13) in the next
form
\begin{eqnarray}
 {\bf G}_{ret} & = G_{ret}+ G_{ret}\circ M_{ret}\circ {\bf G}_{ret} \\
 {\bf G}_{adv} & = G_{adv}+ G_{adv}\circ M_{adv}\circ {\bf G}_{adv}  \\
 {\bf G}_{1} & = G_{1}+ G_{ret}\circ M_{ret}\circ {\bf G}_{1}+
G_{1}\circ M_{adv}\circ {\bf G}_{adv}+G_{ret}\circ M_{1}\circ {\bf G}_{adv}
\end{eqnarray}

\bigskip
{\underline{\it 3.4. Formal solution of the integral equations.}}
\bigskip

   Having derived Eq.(3.27) one usually follows the next way [6]. First he acts
on this equation separately from the left and from the right by the
differential operator of free wave equation  and takes the difference of the
new
equations. Next he transforms ${\bf G}_{1}$ to the Wigner variables and splits
(by hands!) the short- and the long-range scales. This procedure results in an
integral-differential "kinetic equation" for the density ${\bf G}_{1}$ in the
phase space. In a short while we will show that the exact equation (3.27) can
be
solved (at least formally) in an extremely remarkable form.

  Let us first introduce the new functions
\begin{eqnarray}
 G_{0}= & G_{ret}-G_{adv}=G_{10}-G_{01} ,  \\
 M_{0}= & M_{ret}-M_{adv}=-M_{10}-M_{01} .
\end{eqnarray}
We may derive from (3.25) and (3.26) that
\begin{equation}
 {\bf G}_{0} = G_{0}+ G_{ret}\circ M_{ret}\circ {\bf G}_{0}+
G_{0}\circ M_{adv}\circ {\bf G}_{adv}+G_{ret}\circ M_{0}\circ {\bf G}_{adv}.
\end{equation}
The sum and the difference of (3.27) and (3.30) give the similar equation for
$G_{10}$ and $G_{01}$
\begin{equation}
 {\bf G}_{01,10} = G_{01,10}+ G_{ret}\circ M_{ret}\circ {\bf G}_{01,10}+
G_{01,10}\circ M_{adv}\circ {\bf G}_{adv} -
G_{ret}\circ M_{01,10}\circ {\bf G}_{adv}  .
\end{equation}
As Eq.(3.25) may be also rewritten in the same form
\begin{equation}
 {\bf G}_{ret} = G_{ret}+ G_{ret}\circ M_{adv}\circ {\bf G}_{adv}+
G_{ret}\circ M_{ret}\circ {\bf G}_{ret} -
G_{ret}\circ M_{adv}\circ {\bf G}_{adv} ,
\end{equation}
we may use (3.22) and get
\begin{equation}
 {\bf G}_{00,11} = G_{00,11}+ G_{ret}\circ M_{ret}\circ {\bf G}_{00,11}+
G_{00,11}\circ M_{adv}\circ {\bf G}_{adv} +
G_{ret}\circ M_{11,00}\circ {\bf G}_{adv}.
\end{equation}
These routine transformations show that matrix Swinger-Dyson equation do not
mix the different types of Greenians once we recognize the retarded and
advanced functions to play the exclusive role. This role is really singled out.
Indeed, we may rewrite Eq.(3.27) in the form
\begin{equation}
 (1- G_{ret}\circ M_{ret})\circ {\bf G}_{1}=
G_{1}\circ (1+M_{adv}\circ {\bf G}_{adv}) +
G_{ret}\circ M_{1}\circ {\bf G}_{adv}   .
\end{equation}
As ${\bf G}_{ret}\circ M_{ret}\circ G_{ret}={\bf G}_{ret}-G_{ret} $
we easily prove that
\begin{equation}
 (1+ {\bf G}_{ret}\circ M_{ret}) (1- G_{ret}\circ M_{ret})=1.
\end{equation}
Taking into account that
\begin{eqnarray}
 (1+M_{adv}\circ {\bf G}_{adv}) = \rinv \circ {\bf G}_{adv}  ,\nonumber \\
 (1+ {\bf G}_{ret}\circ M_{ret})=   {\bf G}_{ret}\circ \linv ,
\end{eqnarray}
where
\begin{equation}
\rinv (x) = i \hat{\partial_{x}}-m,\;\;\;\linv (x) = -i \hat{\partial_{x}}-m,
\end{equation}
and multiplying (3.34) by
$(1+ {\bf G}_{ret}\circ M_{ret}) $ from the left we find:
\begin{equation}
 {\bf G}_{1} = {\bf G}_{ret}\circ  \linv \circ G_{1}
 \circ \rinv \circ {\bf G}_{adv}
+{\bf G}_{ret}\circ M_{1}\circ {\bf G}_{adv}.
\end{equation}
Another equation of this kind are evident:
\begin{eqnarray}
 {\bf G}_{0} & = {\bf G}_{ret}\circ  \linv \circ G_{0}
 \circ \rinv \circ {\bf G}_{adv}
+{\bf G}_{ret}\circ M_{0}\circ {\bf G}_{adv},  \\
 {\bf G}_{10,01} & = {\bf G}_{ret}\circ  \linv \circ G_{10,01}
 \circ \rinv \circ {\bf G}_{adv}
-{\bf G}_{ret}\circ M_{10,01}\circ {\bf G}_{adv},           \\
  {\bf G}_{00,11} & = {\bf G}_{ret}\circ  \linv \circ G_{00,11}
 \circ \rinv \circ {\bf G}_{adv}
+{\bf G}_{ret}\circ M_{11,00}\circ {\bf G}_{adv}.
\end{eqnarray}

The first and the most naive idea is to forget about the arrows and to
rewrite Eqs. (3.38)-(3.40) in the momentum representation. Then in the first
terms of these equations will appear a string like
$(p^{2} - m^{2} )\delta(p^{2}-m^{2} )$ which equals to zero. It reflects that
simple fact that the bare Greenian $G_{1}$ is a solution of the homogeneous
Dirac equation. But this way is inconsistent.  It would immediately close the
back way to a normal perturbative expansion in powers of the coupling constant.

A more attentive look at the Eqs.(3.40)-(3.43)  shows that   the dressed
Greenians ${\bf G}_{AB}$ can be found at least as the formal solution of the
retarded Cauchy problem with bare Greenians  as initial data and self-energies
as the sources. Indeed, having integrated the first items of these equations
twice by parts we find for all the elements of ${\bf G}_{AB}$
\begin{eqnarray}
{\bf G}_{10,01}(x,y)={1 \over 2} \int d \Sigma_{\mu}(\xi)
 d \Sigma_{\nu}(\eta) {\bf G}_{ret}(x,\xi)\gamma^{\mu}
G_{1}(\xi.\eta)\gamma^{\nu} {\bf G}_{adv}(\eta,y) \nonumber \\
-\int d^{4}\xi d^{4} \eta {\bf G}_{ret}(x,\xi)
M_{10,01}(\xi.\eta) {\bf G}_{adv}(\eta,y) , \\
{\bf G}_{00,11}(x,y)= {1 \over 2}\int d \Sigma_{\mu}(\xi)
 d \Sigma_{\nu}(\eta) {\bf G}_{ret}(x,\xi)\gamma^{\mu}
G_{1}(\xi.\eta)\gamma^{\nu} {\bf G}_{adv}(\eta,y) \nonumber \\
-\int d^{4}\xi d^{4} \eta {\bf G}_{ret}(x,\xi)
[\pm G_{0}^{-1}(\xi,\eta)+M_{11,00}(\xi.\eta)] {\bf G}_{adv}(\eta,y),
\end{eqnarray}
In the first terms of (3.42) and (3.43) we took into account that the
integration is going over the space-like hypersurface and that at space-like
$(\xi - \eta)$
\begin{equation}
  G_{AB}(\xi,\eta)=(1/2)G_{1}(\xi,\eta),
  \;\;\; \; A,B=0,1;\;\; (\xi - \eta)^{2}<0.
\end{equation}

   At first glance it is ambiguous to use of $G_{0}^{-1}$ in Eq.(3.43) without
 the arrows. Nevertheless,
  \begin{eqnarray}
{\bf G}_{ret} \circ \rinv \circ {\bf G}_{adv} -{\bf G}_{ret} \circ \linv \circ
{\bf G}_{adv} =
 (1+ {\bf G}_{ret}\circ M_{ret})\circ {\bf G}_{adv}-   \nonumber \\
{\bf G}_{ret}\circ (1+M_{adv}\circ {\bf G}_{adv}) =-{\bf G}_{0}+
{\bf G}_{ret}\circ M_{0}\circ {\bf G}_{adv}=0,
 \end{eqnarray}
 as it follows from (3.42).

   The universal position of the $G_{1}$  in Eqs.(3.42) and (3.43) indicates
that it play a role as the current density of states.

   All the equations from (3.21) to (3.45) remain literally the same after
replacement ${\bf G}_{AB} \rightarrow {\bf D}_{AB}$ and
${\bf M}_{AB} \rightarrow {\cal P}_{AB}$ except
$\gamma^{\mu} d \Sigma_{\mu} \rightarrow  {\stackrel{\leftrightarrow}
{\partial}}^{\mu} d \Sigma_{\mu}$.
Eqs.(3.42) were first derived in [17] in course of the analysis of some
external
field problems in QED (without the radiative corrections).

   The equations (3.42) and (3.43) as well as their copies for boson fields are
the basic ones for the relativistic quantum field kinetics. Only three of them
(for each field) are mutually independent.

\bigskip
{\underline{\it 3.5. Electromagnetic emission from QGP. An example of
the explicit integration.}}
\bigskip

"Solving" Eq.(3.16) for   ${\bf \Delta}_{10}$  in a manner described above we
immediately express the rate of dileptons emission via polarization operator
$\Pi_{10}(-k)$,
\begin{equation}
 k_{1}^{0} k_{2}^{0} {dN_{e^{+}e^{-}}\over d{\bf k}_{1} {d\bf k}_{2}d^{4}x}=
 ie^{2}{{L_{\mu\nu}(k_{1},k_{2})}\over{4(2\pi)^{6}}} {{\Pi^{\mu\nu}_{10}(-k)}
\over {[k^{2}]^{2}} }   ,
\end{equation}
where $[k^{2}]^{-2}$ stands for a product $D_{ret}(k)D_{adv}(k)$
out of the photon mass shell.

   The rate of a photon emission is already expressed in terms of radiation
operator  $\pi^{\mu\nu}_{10}$ which relates to self-energy $\Pi$
by means of an equation,
\begin{equation}
 \pi =\Pi \circ {\bf \Delta} \circ \Delta^{-1}=
\Pi+\Pi \circ {\bf \Delta} \circ \Pi.
\end{equation}
The $\pi_{10}$-component of Eq.(3.47) is as
\begin{equation}
\pi_{10}=\Pi_{10}+\Pi_{10}\circ {\bf \Delta}_{adv}\circ\Pi_{adv}
+\Pi_{ret}\circ {\bf \Delta}_{ret}\circ\Pi_{10}-
\Pi_{ret}\circ {\bf \Delta}_{10}\circ\Pi_{adv}.
\end{equation}
In the approximation used here we neglect all the terms with powers of
$\alpha =e^{2} / 4 \pi $ , greater than 1. So, approximately,
\begin{equation}
\pi_{10}(-k) \approx \Pi_{10}(-k) ,
\end{equation}
and
\begin{equation}
   k^{0}{{dN_{\gamma}}\over{d{\bf k}d^{4}x}} =
   {{ig_{\mu\nu}}\over{2(2\pi)^{3}}} \Pi^{\mu\nu}_{10}(-k).
\end{equation}

\bigskip
{\underline{\it 3.6. The main kinetic  regimes.}}
\bigskip

   The most impressive feature of the equations like  (3.42) and (3.43) is that
the influence of initial data and that of the current interaction are
manifestly
separated. This  allows one to single out the two main kinetic regimes:

 i.{\it Regime of the weak kinetic coupling.}\\

It is applicable to any many-body system where the mean free path
is much greater then the typical screening length. Processes of the emission of
 the heavy dileptons and the photons  or low-mass  dileptons with a high
 energy  are contributed basically by the hard part of the partons spectra and
 should be analysed in frames of this regime.  Collective modes of the hard
 partons are almost indistinctible from the free particles. The elements of
 collective behavior reveal themselves mainly in elementary collision
 processes with a collinear geometry. They result in a cut-off
 defined by the amplitude of the forward scattering or any other physical scale
 which restricts the space-time region of the
 coherent interaction between the partons.

A more formal description of this regime  sounds as follows.
Let us assume that we  found an exact nonperturbative solution of the
many-body problem and a set of the normal modes for this
solution.  Let us compare these modes with the normal modes of the
noninteracting fields which were used to describe the initial statistical
ensemble.  If we find out that for some part of the spectrum of states
the old and the new modes differ  only by small corrections which can be
found perturbatively then this part of the spectrum is in the regime
of the weak coupling.  In the language of the transition processes it means
that  dispersion of the wave packet presenting a free particle in the media
 which is made from the interacting particles is not too strong.  The retarded
propagators of this regime get no essential damping and the phase memory is
kept between the successive interactions. The interaction does not lead to any
new  singularities in the density of states.   In other words it is an
obvious perturbative region of the field theory. The only  difference is
a nontrivial density matrix of the initial state. The perturbation
expansion is  explicitly accounts for the initial data terms in the
Eqs. (3.42) and (3.43).
\bigskip

 ii.{\it Regime of a strong kinetic coupling.}\\

    The regime  of strong kinetic coupling describes that part of the spectrum
where the initial state of a free particles undergo an active
statistical and(or) phase mixing. Its retarded propagators gain the finite
damping which {\it singles out the time arrow}. The memory about initial
correlations is lost during a short time.  The density of states loses its old
singular points and gains the new ones.  When we omit the initial data terms
in the solution  (3.42), (3.43)
of  the field equations (3.13) we make an irreversible step
towards a nonperturbative theory. If we had a reason to neglect the bare
propagators and bare coupling with respect to the exact
self-energies and vertices
then a theory would have possessed  a scale invariance.

If we could single out the strongest part of interaction,  to solve this
problem  nonperturbatively, and describe the solution in terms of the
quasi-particles, then we would come back to a region of  weak coupling with
 respect to the residual interaction.

   The perturbative expansions in the regimes of strong and weak kinetic
coupling much differ.   We will discuss the different versions of perturbative
expansions using the photon and dilepton emission rates as the example.

\renewcommand{\theequation}{4. \arabic{equation}}
\setcounter{equation}{0}
\medskip
{\bf \Large 4.Perturbation theory for quantum field kinetics}\\
\medskip
\bigskip
{\underline{\it 4.1. Skeleton form of the integral equations.}}
\bigskip

  As any reasonable diagram technique do, the Keldysh one, which we have
started from, allows  to assemble certain subsequences of bare diagrams into
large pieces,the irreducible elements of skeleton diagrams, like exact
Greenians, self energies and vertices. Equations (3.13), (3.16), (3.19) and
(3.20) are the results of such a selective resummation in fact. It would be
surprising if we could get anything else as the only tool used to derive these
equations was the Wick theorem and  the presence of additional indices does not
change the topology of the diagrams. So we can reconstruct the perturbation
theory moving in the backward direction from the skeleton expansion of the
exact Schwinger-Dyson equations.

   The quantity which we shall choose to calculate is the observable rate of
emission, expressed via $\Pi^{\mu\nu}_{10}$. Its graph is given at Fig.1
\begin{figure}
\vspace{2.0cm}
\caption{$\Pi_{10}$ dressed by strong interactions}
\end{figure}

   Let us agree about notations.  Thin and bold solid lines depict bare and
dressed quark propagators.  The labeled ends of the line mean $G_{AB}$.
$G_{ret}$ and $G_{adv}$  have the arrows which show the latest argument. Thin
and bold dashed lines will be used for the gluons. The flattened loop with
labeled ends stands  for quark self-energy $M_{AB}$ . The arrow inside a loop
means $M_{ret}$ or $M_{adv}$. The round loops depict gluon self energies
${\cal P}_{AB}$ ,${\cal P}_{ret}$  or  ${\cal P}_{adv}$. The line crossings are
the bare vertices. The dressed vertices are bold.

The skeleton equation for the vertex is given at Fig.2 ,
\begin{figure}
\vspace{3cm}
\caption{Skeleton expansion of $\gamma q\bar{q}$-vertex}
\end{figure}
 The four-fermion vertex ${\bf K}$ is a sum of the graphs of Fig.3.
\begin{figure}
\vspace{3cm}
\caption{Skeleton expansion for vertex $K$}
\end{figure}

The exact Greenians obey the equations given graphically at Fig.4.
\begin{figure}
\vspace{2.5cm}
\caption{Integral equations for ${\bf G}_{AB}$ and  ${\bf G}_{ret,adv}$}
\end{figure}

   These pictures may be easily found in any standard textbook like [18].

   In order to get the perturbative expansion for  $\Pi_{10}$ (Fig.1) one may
start from the skeleton graphs and replace step by step the exact Greenians and
vertices by the iterations arising from the integral equations (3.25)-(3.27)
for
 propagators (Fig.4). The result should not depend upon the order we do this.
So
we have certain freedom to preserve some physically significant features of the
exact solutions.

\bigskip
{\underline{\it 4.2. Perturbation theory in the regime of weak kinetic
coupling.}}
\bigskip

   Let us start first from the formal iterations remembering that the bare
Greenians carry the information about the initial correlations hidden in the
$\rho_{QCD}$.  They are the vacuum ones if the initial state is pure or the
thermal - like if the initial state is the Gibbs ensemble of free fields. In
the
lowest order we restrict ourselves by the set of graphs given at Fig.5.
\begin{figure}
\vspace{4.5cm}
\caption{Low order iterative expansion for ${\bf G}_{AB}$,
 ${\bf G}_{ret,adv}$ and vertex}
\end{figure}

This approximation will be used to calculate the dilepton and photon yield from
QGP in Refs.[8-10].

   Assembling these expansions into $\Pi_{10}$ up to the $\alpha_{s}$
order we get a sum of graphs drawn at Fig.6.
\begin{figure}
\vspace{4.5cm}
\caption{Expansion of $\Pi_{10}$ in the regime of weak kinetic coupling}
\end{figure}

It is nothing but a squared modulus of the  coherent sum of
amplitudes of the real processes of photon emission up to the
$\alpha_{s}$-order. The corresponding graphs are given at Fig.7. Each process
is weighted with the statistical factors of initial correlations.

\begin{figure}
\vspace{3.5cm}
\caption{The processes contributing to dilepton emission in the
$\alpha_{s}$-order in the regime of weak kinetic coupling}
\end{figure}

The first process is the direct annihilation of bare $q \bar{q}$
pair to (virtual) photon, the next three diagrams relate to the same process
with vertex and mass radiative corrections. Then there follow four diagrams
of $q \bar{q}$ -annihilation accompanied  by gluon emission and absorption and
two diagrams of Compton scattering of quark (or antiquark) on gluon with the
photon emission. The first one and the last four loops of Fig.6 are due to real
process while the others are due to vertex and mass corrections. The crossed
lines correspond to the initial correlations.

\bigskip
{\underline{\it 4.3. Perturbation expansion for a strong coupling.}}
\bigskip

   The strong kinetic coupling implies that a memory about initial correlations
is lost. To study this case we shall iterate only the integral equations for
vertex (3.19) (Fig.2) and Bethe-Salpeter kernel $ K$ (Fig.3), keeping the exact
Greenians as the whole. It'll lead us to to the picture presented at  Fig.8.
\begin{figure}
\vspace{4cm}
\caption{Skeleton expansion for $\Pi_{10}$}
\end{figure}

We would have returned to the previous kind of expansion if we had continue the
formal iteration of propagators using equations (3.13)  (Fig.4). But we shall
not do this and perform the manifest splitting of propagators into the parts
originating from initial correlations and the current dynamics given by
equations (3.42), (3.43) and their copies for gluon propagator.

 If we consider the whole system  or some part of its spectrum which is in the
  regime of strong kinetic coupling then we must neglect the initial
correlations and take into account only the current dynamics of quark-gluon
interactions. Let us draw  the first two terms of
expansion of $\Pi_{10}$ in this limit (Fig.9).
\begin{figure}
\vspace{3.5cm}
\caption{Expansion of $\Pi_{10}$ in the limit of strong kinetic coupling}
\end{figure}

    The subsequent kinetic expansion of self-energies is given by  Fig.10:
\begin{figure}
\vspace{3.5cm}
\caption{Expansion of $M_{AB}$ and ${\cal P}_{AB}$
 in the limit of strong kinetic coupling}
\end{figure}

  Remember that the arrows show the later moments of time in retarded and
advanced propagators. So the kinetic diagrams follow the history of the
emitting
system from the past up to the moment of photon emission.

    The more complicated graph relates to the more deep incite to the past
history. Any two new vertices are accompanied by six new retarded propagators
and each of them brings some new damping. So we may hope that the kinetic
perturbation series in the dense matter will converge very rapidly. Physically
this convergence reflects the high rate of statistical mixing.   As the final
answer can not be sensitive to the type of initial correlation we may choose
for
convenience any density matrix of noninteracting  bare quarks and gluons. The
perturbation chain may be safely cut by the calculation of the enough early
loops in one-loop approximation with the bare quark and gluon fields. The only
residue of this choice will be the macroscopic parameters of the density matrix
like $T$, $u^{\mu}$, $\mu $.

We would like to stress that an unusual form (Figs.9,10) of  a perturbation
expansion is nothing but a result of the standard perturbation series
resummation.

\renewcommand{\theequation}{5. \arabic{equation}}
\setcounter{equation}{0}

\medskip
{\bf \Large 5. Renormalization in a presence of statistical ensemble.}
\medskip

   The further extension of the formalism is impossible without explicit
calculations. First, we are to estimate the rate of initial correlations
damping
in order to decide what kind of perturbative expansion is to be chosen. Second,
in course of calculations we shall unavoidably meet the ultraviolet
divergencies.  Let us consider the quark self-energy as an example.   The real
part of $M_{ret}(s)$ is UV-divergent due to the contribution of the vacuum
substate of the statistical ensemble.

So the self-energy should be renormalized. Let us remember first how does look
the residue degree of freedom in the quantum field theory which allows one to
perform the renormalization at all. This freedom exists solely due to the ill
definition of T-ordered products in the S-matrix at coinciding arguments.

   Now we may write down the part of interaction Lagrangian which is
responsible for the one-loop quark self-energy and include the renormalizing
counter term in it
\begin{equation}
    {\cal L}_{int}^{(ren)}(x) =
   g_{0}\bar{q}t^{a}\gamma^{\mu} B^a_{\mu}q +
(Z_{2}-1)[{{i}\over{2}}\bar{q}\gamma^{\mu}
\stackrel{\leftrightarrow}{\partial_{\mu}}q-m\bar{q}q]-\delta m\bar{q}q
\end{equation}
We can repeat all the calculations which have led to Eq.(3.13) and get
\begin{equation}
  {\bf G}_{AB} = G_{AB}+ \sum_{RS} G_{AR}\circ
(M_{RS}+(-1)^{R}\delta_{RS} \Lambda)\circ {\bf G}_{SB}
\end{equation}
with the additional quasilocal term
\begin{equation}
\Lambda(x,y)=-[\delta m+{{1}\over{2}}(Z_{2}-1)(\rinv(x) +\linv(x))]\delta(x-y).
\end{equation}
It is diagonal, real and do not depend upon any parameters of the ensemble as
it
was not averaged with $\rho_{QCD}$.  So it can be naturally used to remove the
vacuum UV-divergence from the real parts of $M_{ret}$ and $M_{adv}$ calculated
in one-loop approximation. The extension of this example to the complete
renormalization scheme is quite evident.  After this step we will get the
finite
theory.  Any  additional renormalization is neither needed nor possible.

  Though we may formally eliminate the UV-divergencies and reduce them to
the quasilocal terms of known form this step doesn't solve the problem of
physical renormalization yet. We are to put forward certain requirements
concerning the values of some observable quantities. In standard field theory
these  are masses of stable   particles in the asymptotic states
on their mass shells. For the quark-gluon plasma such states are obviously
absent. But for that part of the quark spectrum which is  consistent with
the weak kinetic coupling  a standard prescription might be not too
contradictive.

The disappearing of the gap in the fermion density of states is a typical
result of one-loop calculations with equilibrium or nonequilibrium statistical
ensemble of the neutral plasma. This means that the very definition of the
one-particle state with a very low momenta and finite energy at this level of
calculations is ambiguous. So we are  to put forward other requirements
reflecting the new physical situation.

  Any signal from QGP is suitable for this purpose if it has a chance to
be measured in the experiment with the sufficient accuracy.

 Much more attractive looks another choice following from the theoretical
demands. Namely, at very high temperatures the thermodynamic quantities of
any system are defined only by the temperature and the number of degrees of
freedom. They do not depend upon the structure of normal modes or the
magnitude of coupling constant. So the limit of $T\rightarrow \infty$
seems to be the best reference point for determination of
renormalization constants.

\bigskip
{\bf \Large 7. Conclusion.}                \\
\medskip

The paper presents a formalism which allows us to calculate the rates of
such processes as photons and dileptons emission or heavy quarks production in
the quark-gluon plasma.

In more general frames this technique is capable to replace a
standard Feynman approach. Simply, we are to start with the density matrix
generated by the Fock operators from the pure vacuum state of noninteracting
fields. The reader can easily rewrite  some chapters from a standard textbook
on
QED in terms of this approach as an exercise. In some points even more
transparency is reached. E.g., at the tree level the retarded propagators
naturally appear in every place where we usually put them by hands following a
common sense.  The cutting rules and the unitarity come to be a trivial
consequence of the matrix structure: both $S$ and $S^{+}$   contribute any
observable from the very beginning.

For the case of a true thermal equilibrium the kinetic approach also reproduces
everything what can be obtained from the standard Matsubara formalism under
a condition of its consistency. The later is not too restrictive:

i) we are to be sure that the physical interactions in the system are capable
to
maintain the very thermal equilibrium;

ii) if we study the quantities like one-particle distributions the notion of
the
particle's state should be unambiguous.

All the global relations like TFD theorem come to be
trivial consequence of thermal equilibrium and the matrix
structure of the Schwinger_Dyson equations but none of them is
used to define the observables.

The essentially new point is that the theory is not confined to a specific
form of the thermal distributions. Plasma can be out of  thermal and chemical
equilibrium to any extent. Nevertheless we can consider both real and virtual
processes  in a  self-consistent manner.  Usually it is lost in the  computer
simulations of the emission from the nonequilibrium plasma [16,19] - only the
tree level matrix elements are taken into account.

In the next  paper [9] we give an example:  for the dilepton emission from
nonequilibrium plasma a self-consistent account of the radiative corrections
changes the answers with respect to what can be obtained by a naive cut-off of
the infrared singularities.  We found a significant difference with the case of
 equilibrium plasma also.

  This result was obtained using some analytic  approximations for the
nonequilibrium partons distributions.  We found also that an
obviously used scale
$gT$ is not a universal cut-off.

    Nothing can change if we bypass to the totally numerical calculations
using the results of cascade simulations of the nuclear collisions as an input.

In some recent studies [20] the rate of approaching the hot glue to a
thermal equilibrium was estimated. Processes like $gg \leftrightarrow gg$
and $gg \leftrightarrow ggg$  at the tree level were considered.  The
object to be examined to get an answer is the shape of one-particle
distribution. In sequence, the collision term of the kinetic equation
contains integrals over all the external momenta of these reactions except one.
Without a precise account for the radiative corrections against the background
of the nonequilibrium partons distributions we can hardly get a reliable
answer. Hopefully, the field kinetic approach is adequate to this problem.

\medskip
I am indebted to G. Brown, E. Shuryak and the Nuclear theory group at
SUNY at Stony Brook for continuous support.

 I am grateful to A.Abrikosov,Jr., V.Eletsky, I.Ginzburg, L. McLerran,
R.Pisapski, A.Shabad,  E. Shuryak and I.Zahed for
many fruitful and helpful discussions.

\medskip

{\bf REFERENCES}   \\
\medskip
1. T. Matsubara, Progr.Theor.Phys.,14(1955) 351.\\
2. R.P. Feynman, A.R. Hibbs, Quantum mechanics and path integrals,
McGraw-Hill,NY,1965. \\
3. J. Schwinger, Particles, sources and fields, Addison-Wesley, 1970. \\
4. A.A. Abrikosov, L.P. Gor'kov, I.E. Dzyalosinskii, Methods of quantum field
theory in statistical physics (Prentice-Hall, Englewood Cliffs, 1963)   \\
5. Y. Takahashi, H. Umezawa, Collective phenomena, 1975, v.2, p.55. \\
6. L.V. Keldysh, Sov. Phys. JETP 20 (1964) 1018; E.Lifshits, L. Pitaevsky,
Physical kinetics, Nauka, Moscow, 1979.\\
7. N.P. Landsman, Ch.G. van Weert, Phys.Rep. 145 (1987)141;
A. Niemi, G.W. Semenoff, Nucl.Phys. B230 (1984)181.        \\
8. A. Makhlin, Preprint SUNY-NTG-93-10, Stony Brook, 1993\\
9. A. Makhlin, Preprint SUNY-NTG-93-20, Stony Brook, 1993\\
10. A. Makhlin, Preprint SUNY-NTG-93-21, Stony Brook, 1993\\
11. A. Makhlin, JETP 49(1989)238   \\
12. R.Baier, B. Pire, D. Schiff: Phys.Rev D38 (1988)2819;
 T. Altherr, P. Aurenche, T. Becherrawy: Nucl.Phys. B315 (1989)436;
 T. Altherr, P. Aurenche: Z.Phys. C-Particles and Fields 42(1989)99. \\
13. J. J.Kapusta, P. Lichard, D. Seibert: Phys. Rev. D44 (1991) 2774;
 R. Baier et al. Z.Phys.-C 53 (1992) 433.   \\
14. E.Braaten, R. Pisarski: Nucl.Phys. B337 (1990) 569.   \\
15. E. Shuryak, Phys. Rev. Lett. 68 (1992) 3270.             \\
16. For example, K.Geiger: Phys. Rev. D46 (1992) 4965,4986.   \\
17. A. Makhlin, Preprint ITP-84-88E, Kiev, 1984.  \\
18. J.D.Bjorken and S.Drell, Relativistic quantum fields,
 McGraw-Hill,NY,1965.\\
19. P. Lichard, Preprint TPI-MINN-92/51-T, October 1992\\
20. L.Xiong  and E. Shuryak, Preprint SUNY-NTG-93-24.
\end{document}